\PassOptionsToPackage{square, numbers, sort}{natbib}
\documentclass[reprint,aps,pra]{revtex4-2}

\usepackage[utf8]{inputenc}
\usepackage[T1]{fontenc}
\usepackage[english]{babel}

\usepackage[fleqn]{amsmath}
\usepackage{mathtools}
\usepackage{amssymb}
\usepackage{xfrac}

\usepackage{placeins}
\usepackage{graphicx}
\usepackage{caption}
\usepackage[percent]{overpic}

\newcommand{\ii}{\mathrm{i}}  
\newcommand{\ee}{\mathrm{e}}  

\usepackage{xcolor}
\usepackage[normalem]{ulem}

\begin{document}
\bibliographystyle{unsrt}

\title{Quantum Scattering of Spinless Particles in Riemannian Manifolds}
\author{Lars Meschede, Benjamin Schwager, Dominik Schulz, Jamal Berakdar}
\affiliation{Institut für Physik, Martin-Luther-Universität Halle-Wittenberg}
\date{\today}
\begin{abstract}
 Quantum mechanics is sensitive to the geometry of the underlying space. Here, we present a framework for quantum scattering of a non-relativistic particle confined to a two-dimensional space. When the motion manifold hosts localized curvature modulations, scattering occurs from an emergent geometric potential and the metric tensor field. Analytical and full numerical simulations identify the geometric potential as the primary source for low-energy scattering, while the metric tensor field of the curved space governs high-energy diffraction. Compared to flat spaces, important differences in the validity range of perturbation approaches are found and demonstrated by full numerical simulations using combined finite element and boundary element methods. As an illustration, we consider a Gaussian-shaped dent leading to effects known as gravitational lensing. Experimentally, the considered setup is realizable based on geometrically engineered 2D materials.
\end{abstract}
\maketitle

\section{Introduction}
Quantum dynamics under constrained geometries can be approached with different methods. One way is to quantize the constrained classical system, as has been put forward by DeWitt \citep{DeWitt1952, DeWitt1957}, followed by numerous studies (for a review of the development until 1980, we refer to \citep{Marinov1980}). Possible ambiguities in this approach were addressed in Refs. \citep{Mostafazadeh1994, Mostafazadeh1994a}. Another way is to model the constrained quantum dynamics by embedding the constraint in higher-dimensional space which results in a confining potential for the particles \citep{Jensen1971, Costa1981}. This confining potential approach (CPA) entails the use of local coordinate frames attached to the manifold. It leads (for sufficiently strong and homogeneous confinement \citep{Kaplan1997, Liang2022}) to a well-defined decoupling of the motion tangent from that perpendicular to the constraint. The resulting effective tangent Schrödinger equation includes several geometry-induced terms that originate from the embedding procedure \citep{Brandt2017}. The most prominent one is a scalar potential known as the geometric potential.
\par
Analytical and numerical studies revealed several physical phenomena influenced by the geometry of the manifold of motion. For instance, the geometric potential of bumped surfaces induces bound states relevant to quantum dots \citep{Atanasov2007, Silva2013}. Also, electronic transport is sensitive to space geometry \citep{Serafim2021}. A further finding is the modification of waveguide dispersion of surface plasmon polaritons on metallic wires by geometry-induced momenta \citep{Spittel2015}. The dynamics of strain-driven nanostructures may depend strongly on the geometric potential, as well \citep{Ortix2011}. A further study \citep{Shima2009} shows how the physics of the Tomonaga-Luttinger liquids is altered by the curvature of space. The coupling of constrained charged particles to electromagnetic fields was addressed in Refs. \citep{Ferrari2008, Jensen2009, *Jensen2010} and novel spin-orbit coupling mechanisms were proposed \citep{SOC_Curv_2001,SOC_Curv_2013,SOC_Curv_2015,SOC_Curv_2018}.
\par
Experimentally, Szameit \textit{et al.} \citep{Szameit2010} investigated the analog system of a photonic topological crystal and reported good agreement between the measured propagation of light with theoretical simulations based on the effective tangent Schrödinger equation. However, their experiments were reproduced only by properly incorporating the constraining potential. Onoe \textit{et al.} \citep{Onoe2012} observed an alteration of the Luttinger exponent for curved molecules. Further proposed experimental setups include semiconductor-heterostructures \citep{Ando1982}, electrons floating over the surface of liquid helium droplets \citep{Vadakkumbatt2014} or trapped atomic gases (possibly in microgravity) \citep{Zhang2018, Carollo2022}. Especially binary scattering processes of ultracold dipolar atomic gases trapped within flat two-dimensional manifolds have already been studied in some detail, see \citep{Ticknor2011, Koval2014} and references therein.
\par
For investigating geometry-induced effects and testing the quality of the constraining potential, Mostafazadeh \textit{et al.} studied scattering from a localized curvature modulation in a two-dimensional space \citep{Mostafazadeh1996, Oflaz2018, Bui2019}. In this setting, the scattering amplitude is determined by the interplay of the externally applied potential fields, the geometry-induced potential, and the motion in the curved region with deviating metric. In addition, a Gaussian dent was considered analytically using approximation methods that led to closed expressions for the total scattering cross length (which is the two-dimensional analog to the total scattering cross section encountered in three-dimensional studies and should not be confused with the scattering length) and the optical theorem. Following these ideas, Ref. \citep{Anglin2022} compared the classical, quantum mechanical, and semi-classical regimes for a spherical protuberance of a plane.
\par
With this study we wish to extend these investigations beyond the perturbative approaches. Numerical computations support our analysis for different spectrally resolved scattering cross lengths in the elastic channel. The results evidence that the geometric potential plays the dominant role at low energies. For small amplitudes, it is well-captured by first-order perturbation theory, in contrast to the metric tensor field of the curved space whose influence becomes predominant at high energies. Our results demonstrate how dent-like structures in space act as two-dimensional quantum lenses.
\par
The article is organized as follows: In Sec. \ref{sec:Scattering_Theory} we formulate the scattering problem on asymptotically flat two-dimensional Riemannian manifolds according to the CPA and recapitulate and compare results obtained from both the Lippmann-Schwinger equation and the partial wave analysis (PWA). Then, we study a Gaussian dent in Sec. \ref{sec:Gaussian_Dent_Results} and summarize our findings in Sec.\ref{sec:Conclusion}. Finally, derivations of expressions and further analysis of the induced effects are deferred to appendices.

\section{Scattering Theory on a Riemannian Manifold}
\label{sec:Scattering_Theory}
We consider the quasi-free motion of a nonrelativistic spinless particle whose dynamics are constrained to a Riemannian two-dimensional manifold $\left( \mathcal{M}, g \right)$ isometrically embedded in the three-dimensional Euclidean space as $\mathcal{E}: \mathcal{M} \rightarrow \mathbb{R}^{3}$. For simplicity, we restrict ourselves to cases where $\mathcal{M}$ can be globally parameterized by
\begin{align}
    \begin{aligned}
    \mathcal{Y}: \mathbb{R}^{2} \supset Q &\rightarrow \mathcal{M} \subset \mathbb{R}^{3}\ ,\\
    \textbf{q} = (q^{1}, q^{2})^{T} &\mapsto \mathcal{Y}(\textbf{q})\ .
    \end{aligned}
\label{eq:parametrization}
\end{align}
The Riemannian metric tensor field shall be the canonically induced one, and the Levi-Civita connection is assumed to act on the tangent bundle $T\mathcal{M}$.
\par
As stated, there are different ways to formulate quantum mechanics under constraints. The CPA \citep{Jensen1971, Costa1981} is an approach arguably \citep{Costa1981, Maraner1995, Kaplan1997, Schuster2003} close to physical reality. It starts by inserting \eqref{eq:parametrization} together with a normal displacement, so that $\mathcal{X}(\textbf{q},y) = \mathcal{Y}(\textbf{q}) + y\,\textbf{n}$, into the Schrödinger equation and hence pulls it back into the parameter space $Q\subset \mathbb{R}^{2}$. The manifold is introduced to the dynamics by the action of a constraining potential $V_{\lambda}$ that shall possess the following properties in accordance with \citep{Maraner1995, Schuster2003, Jalalzadeh2005, Brandt2017} and references therein: (i) $V_{\lambda}$ depends only on the normal displacement coordinate $y$; (ii) $V_{\lambda}$ has a deep minimum on $\mathcal{M}$ (for $y=0$) so that it can be expanded around $\mathcal{M}$; (iii) $V_{\lambda}$ preserves the gauge group representing the subgroup of the isometry group. This construction enables a mechanism of dimensional reduction. We note that statement (ii) accounts for quantum fluctuations by securing the uncertainty principle of Heisenberg. The CPA thus allows for a perturbative expansion around $\mathcal{M}$ to separate the effective tangent motion from the normal one. The effective tangent Schrödinger equation reads
\begin{align}
    \left[ -\dfrac{h^{2}}{2m_{0}}\,\nabla_{\mu}\nabla^{\mu} + V_{\mathrm{geo}}\,\hat{I} \right]\chi_{\mathrm{t}}(\textbf{q},t) = \ii\hbar\,\chi_{\mathrm{t}}(\textbf{q},t)\ .
\label{eq:tangent_Schroedinger}
\end{align}
The Laplace-Beltrami operator (calling $\vert g\vert \coloneqq \vert \mathrm{det}(g)\vert$) can be expressed as
\begin{align}
    \nabla_{\mu}\nabla^{\mu} = \dfrac{1}{\sqrt{\vert g \vert}}\partial_{\mu}\,\left( \sqrt{\vert g \vert}\,g^{\mu\nu}\,\partial_{\nu} \right)\ ,
\end{align}
and the geometry-induced potential field, in dependence of the Gaussian curvature $K$ and the mean curvature $M$, has the form 
\begin{align}
    V_{\mathrm{geo}} = -\dfrac{\hbar^{2}}{2m_{0}}\,\left( M^{2} - K \right)\ .
\label{eq:geometric_pot}
\end{align}
This potential always acts attractive for the case given by \eqref{eq:parametrization} \citep{Schuster2003}.
\par
Equation \eqref{eq:tangent_Schroedinger} describes the equilibrium states of the particle that is constrained to $(\mathcal{M}, g)$. We proceed in assuming that the scattering occurs from any localized geometric perturbation of a manifold $\mathcal{M}$ that can asymptotically be identified with the Euclidean plane, $\mathcal{M} {\simeq} \mathbb{R}^{2}$, for $\Vert \mathcal{Y}(\textbf{q}) \Vert_{2} \rightarrow \infty$. As seen from the asymptotic metric tensor field, called $\eta$ in what follows, the latter is flat and possesses trivial topology, so we call $\mathcal{M}$ an asymptotically flat manifold. This condition implies that $\mathcal{M}$ exhibits only localized deviations of the flat metric, and the curved region can hence be approximately modeled as having compact support. Thus, the effective Hamilton operator in \eqref{eq:tangent_Schroedinger} can be decomposed in the general form $\hat{H} = \hat{H}^{0} + \hat{H}^{\mathrm{s}}$ with
\begin{widetext}
\begin{align}
    \hat{H}^{0} &= -\dfrac{\hbar^{2}}{2m_{0}}\,\dfrac{1}{\sqrt{\vert\eta\vert}}\partial_{\mu}\,\left( \sqrt{\vert\eta\vert}\,\eta^{\mu\nu}\partial_{\nu} \right) + V_{\mathrm{geo}}^{0}\,\hat{I}\ , \\
    \hat{H}^{\mathrm{s}} &= -\dfrac{\hbar^{2}}{2m_{0}}\,\biggl[ \left( g^{\mu\nu} - \eta^{\mu\nu} \right)\partial_{\mu}\partial_{\nu} + \left( \dfrac{1}{\sqrt{\vert g\vert}}\partial_{\mu}\left( \sqrt{\vert g\vert}g^{\mu\nu} \right) - \dfrac{1}{\sqrt{\vert \eta\vert}}\partial_{\mu}\left( \sqrt{\vert \eta\vert}\,\eta^{\mu\nu} \right) \right)\partial_{\nu} \biggl] + \left[ V_{\mathrm{geo}} - V_{\mathrm{geo}}^{0} \right]\,\hat{I}\ .
\label{eq:Hamilton_scattering}
\end{align}
\end{widetext}
\noindent
The superscripts $0$ and $\mathrm{s}$ denote the asymptotic and scattered part, respectively. Hence, the origin of the scattering processes are twofold deviations between the asymptotic and scattering regions: The difference of the respective geometry-induced potentials acts as a scalar scattering potential. In addition to that, the disturbance in the metric tensor field yields additional terms that can be regarded as geometry-induced tensor potentials (of first and second rank). The contributions of both terms to the scattering Hamilton operator $\hat{H}^{\mathrm{s}}$ are additive but their effects on scattering observables cannot be disentangled in general. Due to the presence of the derivative terms, the scattering processes investigated here obtain a stronger $\textbf{k}$-dependence than those found in the familiar situation of pure scalar potential scattering. It is this influence that we will identify to be responsible for our observations. $\hat{H}^{0}$ governs the asymptotically free particle according to the eigenvalue problem
\begin{align}
    \hat{H}^{0} \chi_{\mathrm{t}}^{0}(\textbf{q}) = E\,\chi_{\mathrm{t}}^{0}(\textbf{q}) \ .
\label{eq:asymp_EVP}
\end{align}
\par
In the following part, we will study Riemannian manifolds that are globally parameterized in Monge form with
\begin{align*}
\mathcal{Y}(\textbf{q}) \coloneqq \left(q^{1}, q^{2}, f(q^{1}, q^{2})\right)^{T}\ ,
\end{align*}
referring to standard Cartesian coordinates. That means $Q = \mathbb{R}^{2}$ can be identified so that $q^{1} = x$, $q^{2} = y$, and $z=f(x,y)$ with $f$ is assumed to be a smooth, axially symmetric function with vanishing radial derivative at its center located at a certain $\textbf{q}_{0}$. Furthermore, it is expected to decay to a constant value so that
\begin{align*}
    f(\textbf{q}) \simeq z_{0} \in\mathbb{R}\ ,\ \left( \Vert \textbf{q} \Vert_{2} \rightarrow \infty \right)\ ,
\end{align*}
identifying the resulting $\mathcal{M}$ as asymptotically flat. Then, using the reparametrization with respect to polar coordinates, fulfilling the relations $r = \sqrt{(x-x_{0})^{2} + (y - y_{0})^{2}}$ and $\mathrm{tan}(\varphi) = \frac{y-y_{0}}{x - x_{0}}$ so that $f=f(r)$, because of the axial symmetry, we find the Riemannian metric tensor field
\begin{align}
    \left(g_{\mu\nu} \right) = \left( \begin{matrix}
    1 + (\partial_{r}f)^{2} & 0 \\
    0 & r^{2}
    \end{matrix} \right)\ .
\label{eq:Monge_i}
\end{align}
From this, we deduce the Gaussian and mean curvatures,
\begin{align}
    K = \dfrac{F\,\partial_{r}F}{r}\ ,\quad M = \dfrac{1}{2}\,\left( \dfrac{F}{r} + \partial_{r}F \right)\ , 
\label{eq:Monge_pot}
\end{align}
and, consequently, the geometric potential \eqref{eq:geometric_pot},
\begin{align}
    V_{\mathrm{geo}} = -\dfrac{\hbar^{2}}{8m_{0}}\,\left( \dfrac{F}{r} - \partial_{r}F \right)^{2}
    \label{eq:geom_Pot_Monge}
\end{align}
with the definition 
\begin{align}
    F \coloneqq \dfrac{\partial_{r}f}{\sqrt{1 + (\partial_{r}f)^{2}}}\ .
\label{eq:Monge_f}
\end{align}

\subsection{Effective Tangent Lippmann-Schwinger Equation}
For studying scattering processes, we use the Lippmann-Schwinger equation which reads
\begin{align}
    \vert \chi_{\mathrm{t}}^{\pm} \rangle = \vert \chi_{\mathrm{t}}^{0} \rangle + \left[ (E \pm \ii\varepsilon)\,\hat{I} - \hat{H}^{0} \right]^{-1} \hat{H}^{\mathrm{s}}\vert \chi_{\mathrm{t}}^{\pm}\rangle
\end{align}
for the state vectors ($+$ or $-$ correspond to outgoing or incoming wave boundary conditions). In the representation parameterized by \eqref{eq:parametrization} ($\langle \textbf{q} \vert\equiv\langle \textbf{x}(\textbf{q}) \vert$), we find for the wave function
\begin{align}
\begin{aligned}
    \langle \textbf{q} \vert \chi_{\mathrm{t}}^{\pm} \rangle &= \langle \textbf{q} \vert \chi_{\mathrm{t}}^{0} \rangle + \dfrac{2m_{0}}{\hbar^{2}}\,\int G^{\pm}(\textbf{q},\textbf{q}')\,\langle \textbf{q}' \vert \hat{H}^{\mathrm{s}} \vert \chi_{\mathrm{t}}^{\pm} \rangle\ \mathrm{d}\sigma(\textbf{q}') \\
    &= \langle \textbf{q} \vert \chi_{\mathrm{t}}^{0} \rangle + \langle \textbf{q} \vert \chi_{\mathrm{t}}^{\mathrm{s}\,\pm} \rangle\ ,
\end{aligned}
\label{eq:wave_profile_abstract}
\end{align}
where
\begin{align}
    G^{\pm}(\textbf{q},\textbf{q}') = \dfrac{\hbar^{2}}{2m_{0}}\,\left\langle \textbf{q} \biggl\vert \left[ (E \pm \ii\varepsilon)\,\hat{I} - \hat{H}_{0} \right]^{-1} \biggl\vert \textbf{q}'\right\rangle
\end{align}
is a Green's function for the asymptotic form of \eqref{eq:tangent_Schroedinger}, i.\,e., $\hat{H}\simeq\hat{H}^{0}$.
\par
Expression \eqref{eq:wave_profile_abstract} describes the stationary spatial profile of the wave function due to the scattering event. It consists of a superposition of the incident wave and the scattered wave labeled by the superscripts $0$ and $\mathrm{s}$, respectively. We identify
\begin{align*}
    \langle \textbf{q} \vert \chi_{\mathrm{t}}^{\mathrm{s}\,\pm} \rangle &= \dfrac{2m_{0}}{\hbar^{2}}\,\int G^{\pm}(\textbf{q},\textbf{q}')\,\langle \textbf{q}' \vert \hat{H}^{\mathrm{s}} \vert \chi_{\mathrm{t}}^{\pm} \rangle\ \mathrm{d}\sigma(\textbf{q}') \\
    &\sim a(\textbf{k},\textbf{k}^{\mathrm{s}})\ ,
\end{align*}
where $a$ is the scattering amplitude between states characterized by the wave vectors $\textbf{k}$ and $\textbf{k}^{\mathrm{s}}$. In the case of asymptotically flat manifolds, the asymptotic form of the effective tangent Schrödinger equation \eqref{eq:tangent_Schroedinger} is the two-dimensional Helmholtz equation. This possesses plane wave solutions so that $\langle\textbf{q} \vert \chi_{\mathrm{t}}^{0} \rangle = \frac{1}{2\pi}\,\ee^{\ii\textbf{k}\cdot\textbf{q}}$. From \eqref{eq:wave_profile_abstract}, we derive the wave profile incorporating the outgoing waves
\begin{align}
    \langle \textbf{q}\vert \chi_{\mathrm{t}}^{+} \rangle \simeq \dfrac{1}{2\pi}\,\left( \ee^{\ii\textbf{k}\cdot \textbf{q}} + a(\textbf{k},\textbf{k}^{\mathrm{s}})\,\dfrac{\ee^{\ii kr}}{\sqrt{r}} \right)\ .
\label{eq:wave_profile}
\end{align}
The scattering amplitude is found to be
\begin{align}
    a(\textbf{k},\textbf{k}^{\mathrm{s}}) = \dfrac{m_{0}\,\ee^{-\ii\frac{3\pi}{4}}}{\hbar^{2}}\,\sqrt{\dfrac{2\pi}{k}}\,\int \ee^{-\ii\textbf{k}^{\mathrm{s}}\cdot \textbf{q}'}\,\langle \textbf{q}' \vert \hat{H}^{\mathrm{s}} \vert \chi_{\mathrm{t}}^{+} \rangle\ \mathrm{d}\sigma(\textbf{q}')
\label{eq:scat_amp_LS}
\end{align}
which is obtained by inserting the asymptotic form of the Green's function \citep{Sakurai2020} that is proportional to the Hankel function of the first kind \citep{Standards2010},
\begin{align*}
    G(\textbf{q}, \textbf{q}') &= -\dfrac{\ii}{4}\,H^{(1)}_{0}(k\,\Vert \textbf{q} - \textbf{q}' \Vert_{2}) \\
    &\simeq -\dfrac{\ii}{4}\,\sqrt{\dfrac{2}{\pi k r}}\,\ee^{\ii\,\left( kr + \textbf{k}^{\mathrm{s}} \cdot \textbf{q}' - \frac{\pi}{4} \right)}\ ,\ \left( r \rightarrow \infty \right)\ ,
\end{align*}
with $k=\Vert \textbf{k}\Vert_{2}$, $r=\Vert \textbf{q}\Vert_{2}$, and $\textbf{k}^{\mathrm{s}} = k\,\textbf{e}_{r}$. Setting the incident direction without loss of generality as the positive $x$-direction, $\textbf{k} = k\,\textbf{e}_{x}$, we can express the scattering amplitude in dependence of the scattering angle $\vartheta\in [0,\pi]$ between the incident and scattered wave vectors $\textbf{k}$ and $\textbf{k}^{\mathrm{s}}$. Thus, the differential scattering cross length is given by $\frac{\mathrm{d}\sigma}{\mathrm{d}\vartheta} \coloneqq \vert a(\vartheta) \vert^{2}$. The total scattering cross length and the momentum transfer cross length are
\begin{align}
    \sigma_{\mathrm{tot}} = \int\limits_{0}^{2\pi} \vert a(\vartheta) \vert^{2}\ \mathrm{d}\vartheta
\label{eq:total_cross_length}
\end{align}
and
\begin{align}
    \sigma_{\mathrm{M}} = \int\limits_{0}^{2\pi} \left( 1- \mathrm{cos}(\vartheta) \right)\,\vert a(\vartheta) \vert^{2}\ \mathrm{d}\vartheta\ .
\label{eq:momtrans_cross_length}
\end{align}
Normalizing the partial cross length to the total one, we obtain a quantity that can be interpreted as a probability density,
\begin{align}
    w(\vartheta) \coloneqq \dfrac{1}{\sigma_{\mathrm{tot}}}\,\dfrac{\mathrm{d}\sigma}{\mathrm{d}\vartheta}(\vartheta)\ ,
\end{align}
so that $\mathrm{d}p(\vartheta) = w(\vartheta)\,\mathrm{d}\vartheta$ describes the probability of an incident wave being scattered into the infinitesimal angular segment around a certain direction.
\par
By inserting the results of the Monge parametrization \eqref{eq:Monge_i} -- \eqref{eq:Monge_f}, we obtain, as in Ref. \citep{Oflaz2018}, an expression for the scattering amplitude in the first Born approximation, $a(\textbf{k},\textbf{k}^{\mathrm{s}}) \simeq a^{(1)}(\textbf{k},\textbf{k}^{\mathrm{s}})$ (full details are in the appendix),
\begin{widetext}
\begin{align}
\begin{aligned}
    a^{(1)}(\vartheta) = \ee^{\ii\frac{\pi}{4}}\,\sqrt{\dfrac{\pi}{2k}}\,\int\limits_{0}^{\infty} r'\,&\left[ \left( F^{2}k^{2}\,\mathrm{sin}^{2}\left( \sfrac{\vartheta}{2} \right) + \dfrac{1}{4}\,\left( \dfrac{F}{r'} - \partial_{r'}F \right)^{2} \right)\,J_{0}(2\,k\,r'\,\mathrm{sin}\left( \sfrac{\vartheta}{2} \right)) \right.\\
    & \hspace*{3mm}\left. + \left( \dfrac{F^{2} k}{2\,r'\,\mathrm{sin}\left( \sfrac{\vartheta}{2} \right)} + k\,\mathrm{sin}\left( \sfrac{\vartheta}{2} \right)\,F\,\partial_{r'}F \right)\,J_{1}(2\,k\,r'\,\mathrm{sin}\left( \sfrac{\vartheta}{2} \right)) \right]\ \mathrm{d}r'\ ,
\end{aligned}
\label{eq:1B_scat_amp_main}
\end{align}
\end{widetext}
\noindent
where $J_{n}$ are Bessel functions of the first kind of order $n\in\mathbb{N}_{0}$. We want to emphasize that the formula \eqref{eq:1B_scat_amp_main} is valid for general axially symmetric asymptotically flat two-dimensional submanifolds of Euclidean space $\mathbb{R}^{3}$. Its integrand consists of a sum of four contributions. One of them (the second one in the first row) is given by the geometric potential field \eqref{eq:geom_Pot_Monge}, and the remaining terms depend directly on the deformation via the quantity $F$. These other contributions confirm that the non-flat metric acts as a scattering center itself, thus generalizing the short-range potential scattering process.
\par
To elucidate the influence of these terms on the first Born scattering amplitude, we compare the familiar situation of short-ranged potential scattering \citep{Lapidus1982} in flat space where only an external potential may be present as an additional term in the integrand of \eqref{eq:1B_scat_amp_main}. There, we would have $F=0$ and hence no geometric potential, leading to the vanishing $a^{(1)} \rightarrow 0$ as $k\rightarrow \infty$. Typically, the first Born approximation yields a reliable description of this energy regime. For the scattering upon geometric structures, however, it behaves differently because the wave has to travel through a region with nontrivial metric. We can infer that the geometric potential obtains a role similar to that of an external scattering potential. Inspecting the spectrally resolved scattering amplitude, one deduces from the explicit and implicit (in the argument of the Bessel functions) dependencies on $k$ that the geometric potential dominates the process in the regime of small incident energies, leading to the known divergent behavior of the total scattering cross length, $\sigma_{\mathrm{tot}}^{(1)}\sim k^{-1}$, as it is attractive \citep{Papoular1985}. In the limit of large energies, its influence becomes marginal so that the remaining terms determine the scattering process in \eqref{eq:1B_scat_amp_main}.

\subsection{Effective Tangent Partial Wave Analysis}
The PWA is a standard approach in scattering theory. We refer, for example, to \citep{Lapidus1982, Lapidus1986} for flat two-dimensional spaces, i.\,e. $\mathcal{M} = \mathbb{R}^{2}$, and to \citep{Burke2011} for the three-dimensional analog. Performing it, using the ansatz \eqref{eq:wave_profile} for the wave profile, we express the scattering amplitude as (see appendix)
\begin{align}
    a(\vartheta) = \ee^{-\ii\frac{\pi}{4}}\sqrt{\dfrac{2\pi}{k}} \,\sum\limits_{m=-\infty}^{+\infty} \left( S_{m} - 1 \right)\,\ee^{\ii m\vartheta}\ ,
\end{align}
where $S_{m} \coloneqq \ee^{2\ii\delta_{m}}$ is the S-matrix of the corresponding orbital angular momentum channel in dependence of the scattering phase shift $\delta_{m}$. With this expression, the cross lengths \eqref{eq:total_cross_length} and \eqref{eq:momtrans_cross_length} attain the forms
\begin{align}
\begin{aligned}
    \sigma_{\mathrm{tot}} &= \sum\limits_{m= -\infty}^{+\infty}\sigma_{m} = \dfrac{1}{k}\,\sum\limits_{m=-\infty}^{+\infty} \left\vert S_{m} - 1 \right\vert^{2} \\
    &= \dfrac{4}{k}\,\sum\limits_{m=-\infty}^{+\infty} \mathrm{sin}^{2}(\delta_{m})\ ,
\end{aligned}
\label{eq:total_cross_length_PWA}
\end{align}
where $\sigma_{m}$ is the partial scattering cross length of the respective orbital angular momentum channel with $m\in\mathbb{Z}$, and
\begin{align}
\begin{aligned}
    \sigma_{\mathrm{M}} &= \dfrac{1}{2k}\,\sum\limits_{m=-\infty}^{+\infty} \left\vert S_{m+1} - S_{m} \right\vert^{2} \\
    &= \dfrac{2}{k}\,\sum\limits_{m=-\infty}^{+\infty} \mathrm{sin}^{2}(\delta_{m+1} - \delta_{m})\ .
\end{aligned}
\label{eq:momtrans_cross_length_PWA}
\end{align}
The relevant data can be derived from the K-matrix as this is the complex Cayley-transform of the S-matrix,
\begin{align}
    K_{m} = \mathrm{tan}(\delta_{m}) = \ii\,\dfrac{1-S_{m}}{1+S_{m}}\ .
\end{align}
For a given spatial profile of the wave function we use the relation
\begin{align}
    K_{m} = \dfrac{\dfrac{p_{m}(r_{1})}{p_{m}(r_{2})}\,J_{m}(k r_{2}) - J_{m}(k r_{1})}{\dfrac{p_{m}(r_{1})}{p_{m}(r_{2})}\,Y_{m}(k r_{2}) - Y_{m}(k r_{1})}\ ,
\label{eq:K_mat_eval}
\end{align}
for computation, where $J_{m}$ and $Y_{m}$ denote the Bessel functions of the first and second kind, respectively, and 
\begin{align}
    p_{m}(r) \coloneqq \left\langle \chi_{\mathrm{t}}(r,\vartheta), \ee^{\ii m\vartheta} \right\rangle_{L^{2}(\mathbb{S}^{1})}
\end{align}
indicates projection of the tangent wave function onto the angular part of the cylindrical harmonics.
\par
To evaluate these formulas, we solve the scattering problem employing a combination of a finite element method and a boundary element method (FEM-BEM) in two dimensions. The two components of our simulations depend on the FEniCSx \citep{Basixpython,Basixpython2,UFLpython} library. After obtaining the wave function $\chi_{\mathrm{t}}$, we compute the K-matrix inserting two arbitrary but fixed values for the radii. From this, we determine both the scattering phase shift and the S-matrix, either of which could be used to yield the cross lengths.

\section{Results for the Gaussian Dent}
\label{sec:Gaussian_Dent_Results}
For illustration, we study a Gaussian dent described by
\begin{align}
f(r) \doteq f_{0}\,\ee^{-\frac{r^{2}}{2\,\sigma^{2}}}\ ,
\label{eq:Gauss_Monge}
\end{align}
with $f_{0}\in\mathbb{R}$ describing the peak amplitude and $\sigma\in\mathbb{R}^{+}$ the standard deviation that act as measures for the depth and width, respectively. Although this function does not have compact support, it decays sufficiently fast so that we can cut it beyond an effective range. Evaluating \eqref{eq:Monge_f} gives
\begin{align}
F = -\dfrac{f_{0}\,r\,\ee^{-\frac{r^{2}}{2\,\sigma^{2}}}}{\sigma^{2}\,\sqrt{1+\left( \frac{f_{0}\,r}{\sigma^{2}} \right)^{2}\,\ee^{-\frac{r^{2}}{\sigma^{2}}}}}\ ,
\label{eq:Gauss_F}
\end{align}
hence the geometric potential \eqref{eq:geometric_pot} reads
\begin{align}
V_{\mathrm{geo}} = -\dfrac{\hbar^2}{8 m_{0}}\cdot \dfrac{f_{0}^{2}\,r^{4}\,\left(f_{0}^{2} + \sigma^{2} \ee^{\frac{r^{2}}{\sigma ^2}} \right)^{2}}{\left( f_{0}^{2}\,r^{2} + \sigma^{4}\,\ee^{\frac{{r}^2}{\sigma^{2}}}\right)^{3}}\ , 
\label{eq:Gauss_Vgeo}
\end{align}
which is axially symmetric, varies radially, and has a ring-shaped structure. A more detailed discussion of the geometric quantities is given in the appendix.
\par
Using FEM-BEM, we solved the direct scattering problem as defined by \eqref{eq:tangent_Schroedinger} numerically and computed the scattering data from the result. For comparison, we repeated the simulation twice, neglecting either the deviation of the metric or the geometric potential, which allows for estimating the individual contributions to the geometric scattering process.

\subsection{Discussion of the Scattering Data for Varying Geometric Parameters}
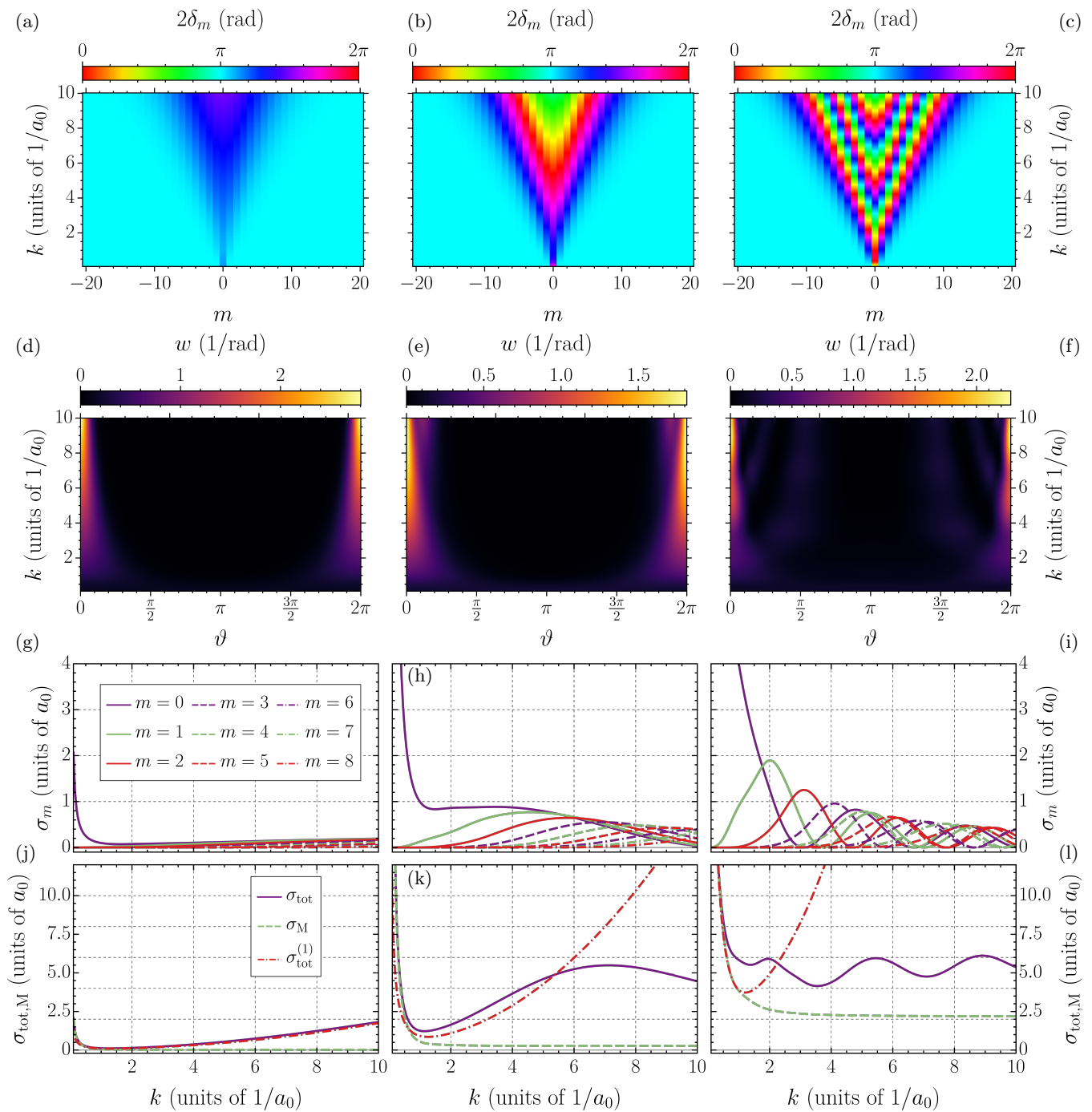
\begin{figure*}
\centering
\begin{overpic}[width=\textwidth]{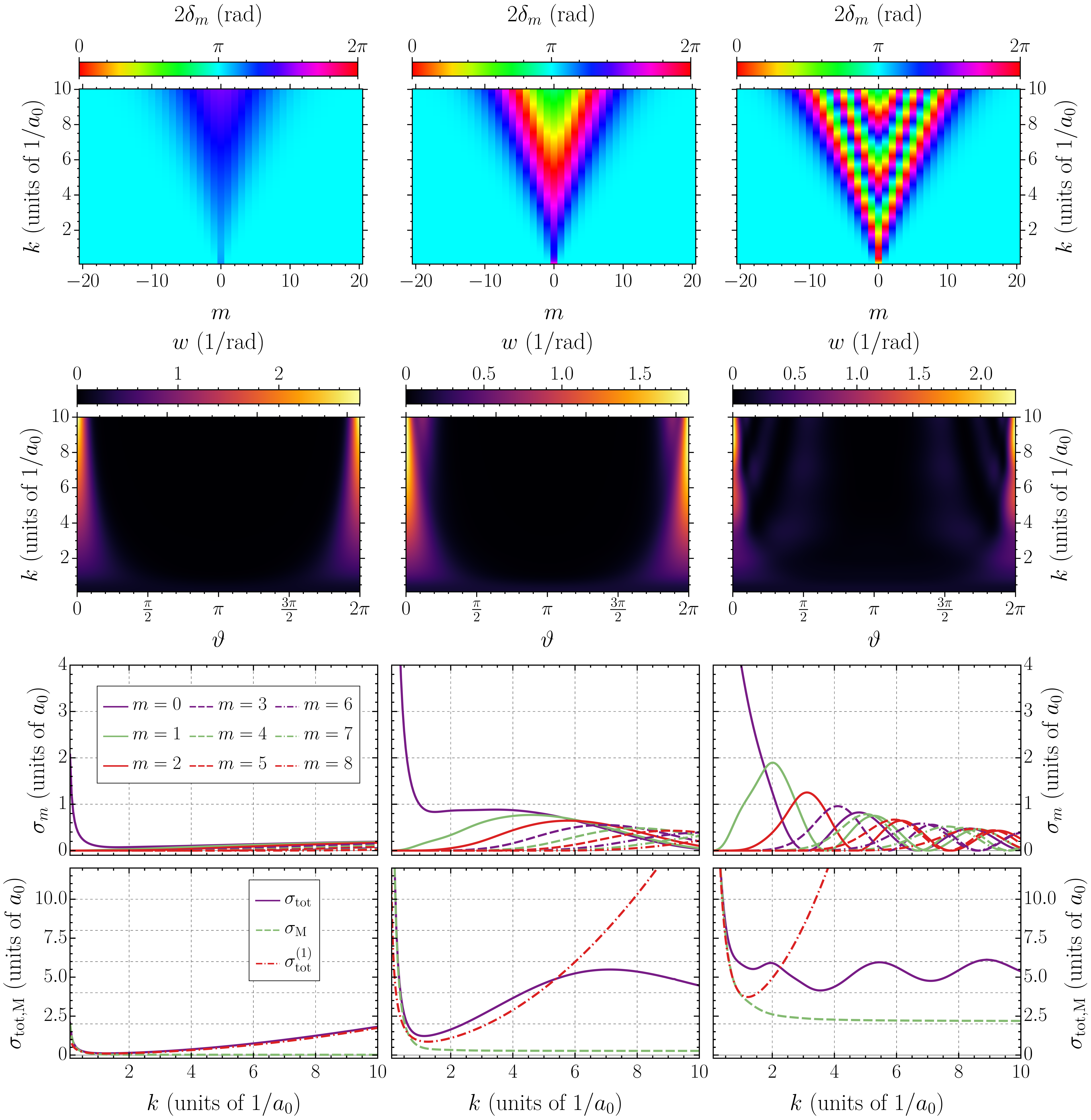}
\put(1,98){(a)}
\put(1,68.7){(d)}
\put(1,42){(g)}
\put(1,23){(j)}
\put(36.3,98){(b)}
\put(36.3,68.7){(e)}
\put(36.3,39){(h)}
\put(36.3,20.7){(k)}
\put(95,98){(c)}
\put(95,68.7){(f)}
\put(95,42){(i)}
\put(95,23){(l)}
\end{overpic}
\caption{(a-c) Directional scattering probability amplitude as function of the scattering angle; (d-f) scattering phase shifts for the first few orbital angular momentum channels; (g-i) partial scattering cross lengths in dependence of the incident wavenumber; (j-l) total and momentum transfer scattering cross lengths in dependence of the incident wavenumber for the direct scattering problem upon a Gaussian dent. Overall, $\sigma = a_{0}/\sqrt{2}$ and $f_{0} \in \{0.5; 1; 2\}\,a_{0}$ in the columns from left to right. A characteristic length of the problem serves as the unit $a_{0}$.}
\label{fig:scat_data_w_delta}
\end{figure*}
Fig.\,\ref{fig:scat_data_w_delta} presents results for the scattering process. The two lower rows of fig.\,\ref{fig:scat_data_w_delta}\,(g)-(l) show the partial, total, and momentum transfer cross lengths obtained from both the analytical approach using the first Born approximation and our numerical simulations. The two upper rows in fig.\,\ref{fig:scat_data_w_delta}\,(a)-(f) show the corresponding energy resolved directional scattering probability density $w$ in dependence of the scattering angle $\vartheta$ and the scattering phases $\delta_{m}$ for all orbital angular momentum channels. Different amplitudes $f_{0}\in\{ 0.5; 1; 2 \}\,a_{0}$ of the dent are associated with the columns, where $\sigma = a_{0}/\sqrt{2}$ is kept constant and a unit of $a_{0} = 10\,\mathrm{nm}$ was chosen. Higher amplitudes determine larger variation in the metric and hence stronger curvature.
\par
The integrated cross lengths evidence that localized geometric deviations cause scattering processes. In order to understand them, at first, we can compare the results for the spectrally resolved total cross length given by the two approaches (compare in fig.\,\ref{fig:scat_data_w_delta}\,(j)-(l) the purple solid line (full numerics) with the red dot-dashed line (analytical first Born approximation)). Thereby, we recognize significant deviations, which are the stronger, the greater the deformation of space is. Particularly at higher energies (roughly for $k\gtrsim 1\,a_{0}^{-1}$), the first Born approximation appears to be not sufficient to describe the simulation. Better agreement is found for low wave vectors. According to \eqref{eq:1B_scat_amp_main} and the discussion of the significance of its different contributions in the low- and high-energy regimes, we can firstly conclude that the scattering process upon localized geometric structures is indeed governed by the scalar geometric potential in the low-energy regime, stating a familiar situation for which the first Born approximation yields a reasonable description. Secondly, it expects leading contributions by the metric tensor field for higher energies but does not capture these properly. However, the known behavior is reproduced in the limiting case of flat spaces. Based on the first Born approximation, scattering from localized geometric structures was proposed as a possibility to study geometry-induced potential fields \citep{Mostafazadeh1996, Oflaz2018, Bui2019}.
\par
At higher values of $k$, the numerical results reveal an oscillatory behavior of the total cross length while the momentum transfer cross length (see the green dashed line in fig.\,\ref{fig:scat_data_w_delta}\,(j)-(l)) converges. Interpreting these as measures of the lateral size of the target, we infer that the momentum transfer cross length delivers a systematically lower result than the total one. In addition, both increase as the degree of deformation of space increases, which means more scattering. That is, higher amplitudes $f_{0}$ of the dent and hence its overall shape affects the quality of the scattering cross lengths as size measures in the observed spectral range because the width parameter $\sigma$ was kept constant in all three cases. This behavior is in contrast to pure scalar potential scattering and is caused by the tensor potentials.
\par
Considering the scattering probability densities in figs.\,\ref{fig:scat_data_w_delta}\,(d)-(f), we notice two scattering characteristics: while the pattern is isotropic for low energies, anisotropic behavior with a pronounced forward peak is visible for high ones. The partial cross lengths depicted in figs.\,\ref{fig:scat_data_w_delta}\,(g)-(i) confirm that the low-energy process is governed by the $\mathrm{s}$-channel ($m=0$). Indeed, this is the only one yielding a nonvanishing component of the total scattering cross length and a nonzero scattering phase shift in the relevant regime (see figs.\,\ref{fig:scat_data_w_delta}\,(a)-(c)). From the scattering phase shifts follows that the number of involved scattering channels grows linearly with the incident wavenumber $k$ and that channels with opposite quantum numbers of orbital angular momentum contribute symmetrically (${S_{m} = S_{-m}\ \forall\ m\in\mathbb{Z}}$). Generally, the phases become more sensitive to the incident wavenumber when the curvature of the target varies stronger.
\par
As a consequence, an interference pattern emerges within the wave profile that can also be observed in the angular dependence of the scattering probability densities for higher incident wavenumbers. This effect can be understood from a wave mechanical point of view (see below). Recalling that the geometric potential is less critical for the regime of high incident energies, the target is essentially reduced to a curved region through which the wave travels freely. Then, there is a family of elementary waves that pass the structure along continuously varying geodesic trajectories. As these possess different path lengths, the respective representatives of the family experience varying phase shifts while traversing the target, and their recombination behind it leads to the interference effect. Therefore, scattering of an incoming plane wave upon a Gaussian dent leads to diffraction, and the substructure visible in the density plots figs.\,\ref{fig:scat_data_w_delta}\,(d)-(f) shows the diffraction orders. Their number increases with the incident wavenumber.

\subsection{Influences of the Metric and the Geometric Potential}
From \eqref{eq:Hamilton_scattering} follows that scattering is due to both variation in the metric and the induced geometric potential. Here, we study these two effects separately. To this end, we use the parameters $f_{0} = a_0$ and $\sigma = a_0/\sqrt{2}$.
\par
\begin{figure*}[!]
\begin{overpic}[width=0.67\textwidth]{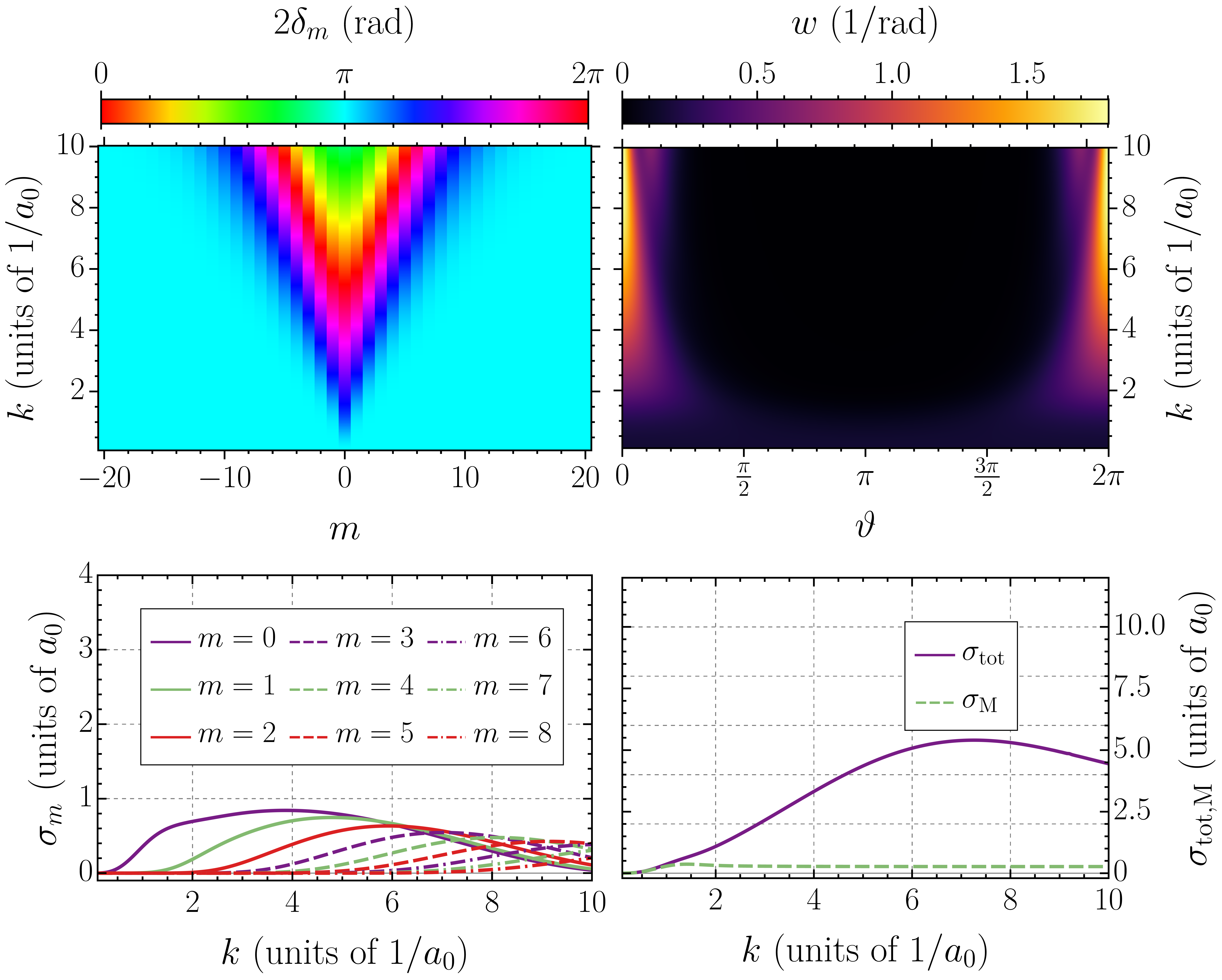}
\put(1,70){(a)}
\put(95,70){(b)}
\put(1,35){(c)}
\put(95,35){(d)}
\end{overpic}
\caption{Directional scattering probability (a), scattering phase shifts (b), partial (c), and integrated scattering cross lengths for the scattering problem without the geometric potential $V_{\mathrm{geo}}$.}
\label{fig:wo_Vgeo}
\end{figure*}

Fig.\,\ref{fig:wo_Vgeo} shows the results when the geometric potential is not taken into account. Thus, the projectile travels freely but through a curved region. In the regime of high energies, the results of all plots match those of the above full analysis (fig.\,\ref{fig:scat_data_w_delta}), which is expected as the contribution of the geometric potential vanishes there. For small energies ($k\lesssim a_{0}^{-1}$), the disappearance of the attractive geometric potential leads to a vanishing of all partial and integrated scattering cross lengths. Hence, we confirm that the geometric potential drastically influences the low-energy scattering and is responsible for the previously observed divergences of scattering cross lengths.
\par
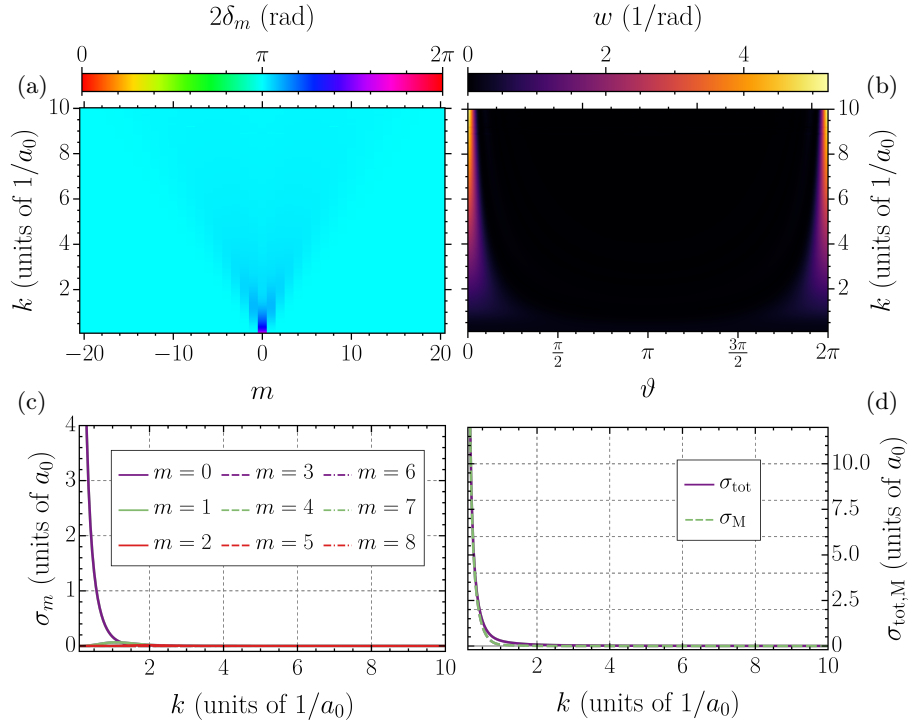
\begin{figure*}[!]
\begin{overpic}[width=0.67\textwidth]{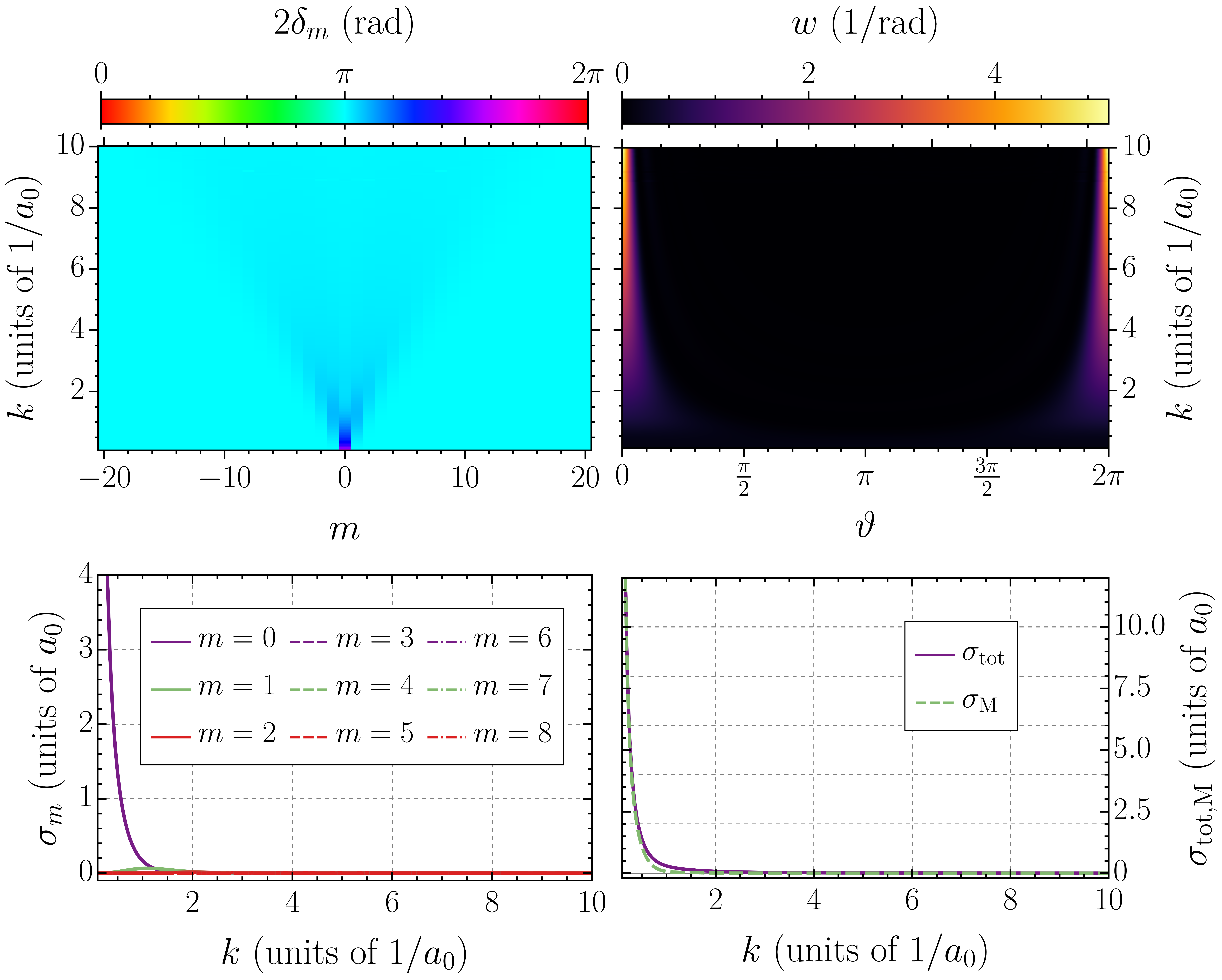}
\put(1,70){(a)}
\put(95,70){(b)}
\put(1,35){(c)}
\put(95,35){(d)}
\end{overpic}
\caption{Directional scattering probability (a), scattering phase shifts (b), partial (c), and integrated scattering cross lengths for the scattering problem with the metric of a flat space.}
\label{fig:wo_metric}
\end{figure*}

Fig.\,\ref{fig:wo_metric} presents the results for scattering in flat space off an external potential field that has the form of our geometric potential \eqref{eq:Gauss_Vgeo}. Here, we observe the typical divergence for small wavenumbers and a scattering cross length approaching zero for high ones, where the scattering from the potential is negligible. Thus, the deviation of the metric yields significant contributions to scattering processes that originate from the different geodesic pathways of the elementary waves and dominate the high energy regime. The expected scattered wave profiles need to be fundamentally changed compared to the flat space scenario in order to produce the observed interference effect. Scattering processes thus appear sensitive to the structure of the underlying space.

\subsection{Lensing by Gaussian Dent}
\begin{figure}[!]
\includegraphics[width=0.34\textwidth]{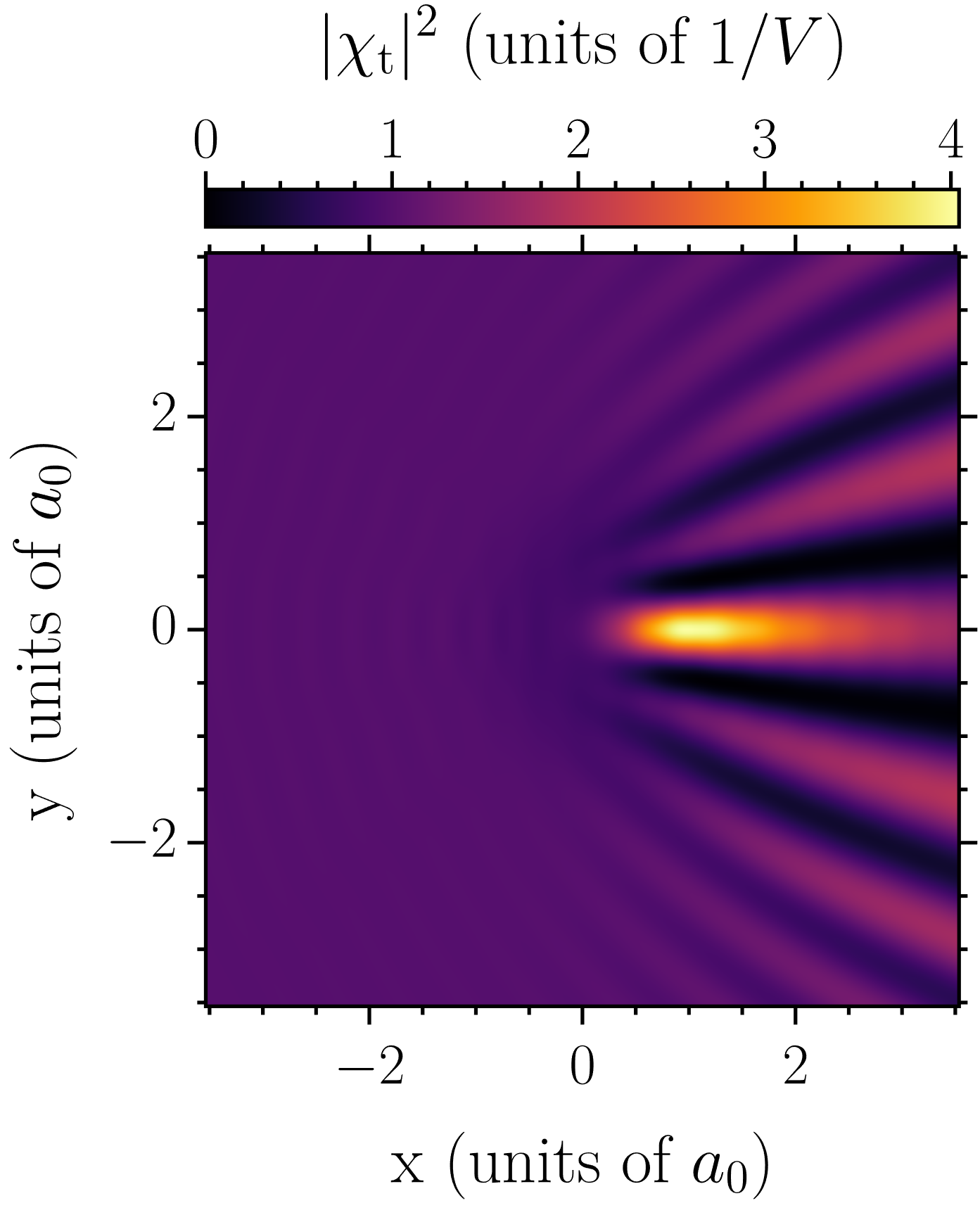}
\caption{Absolute square of the wave function at $k=7.5\,a_0^{-1}$, $f_0=a_0$, and $\sigma = a_{0} / \sqrt{2}$.}
\label{fig:focus_abs_psi}
\end{figure}

Fig.\,\ref{fig:focus_abs_psi} shows the squared absolute value of the amplitude of the stationary wave profile according to equation \eqref{eq:wave_profile}, illustrating how the Gaussian dent acts like a lens for the two-dimensional plane wave $\chi_{\mathrm{t}}^{0}\propto \ee^{\ii k x}$ incoming from $x\rightarrow -\infty$. We observe an alteration of the probability density in the form of a diffraction pattern with a pronounced focus centered around the forward direction. The effect is found for different incident wavenumbers, with the point of maximal probability amplitude shifting along the positive $x$-axis as the energy of the projectile is increased.
\par
Assuming that the wavelength $\lambda = 2\pi/k$ is small compared to the expansion of the dent, which holds for high energies, we can describe this phenomenon by elementary waves that travel through the curved region, accumulating a phase shift due to path elongation (comparing the arc lengths of the (classical) geodesic trajectories they follow). As shown in the previous section, the influence of the geometric potential is marginal for large incidence wavenumbers $k$ as $\mathrm{max}\{V_{\mathrm{geo}}\} \ll \frac{\hbar^{2}}{2 m_{0}} k^2$. Therefore, the oscillating behavior of the total scattering cross length (especially seen for large $f_0$, compare fig.\,\ref{fig:scat_data_w_delta}\,(l)) is caused by the change of the metric tensor and can be qualitatively explained by an approximate one-dimensional model, as follows.
\par
Neglecting the geometric potential and the momentum in $y$-direction, the approximate differential equation (see \ref{eq:tangent_Schroedinger}) reads
\begin{align}
    \left[ \frac{1}{\sqrt{1 + (\partial_x f)^2}}\partial_x\left(\frac{1}{\sqrt{1 + (\partial_x f)^2}}\partial_x \right) + k^2\,\hat{I} \right] \chi_{\mathrm{t}} = 0 \ .
\end{align}
Therefore, its general solutions are given by
\begin{align}
    \chi_{\mathrm{t}}(x) = c_1 \sin\!\left(c_2 + k s(x) \right) \ ,
\end{align}
with $c_1, c_2 \in \mathbb{C}$ and a phase change proportional to the arc length $s(x) \coloneqq \int_{x_0}^{x} \sqrt{1 + (\partial_{x'} f)^2}\ \mathrm{d}x'$ of the curve within $(\mathcal{M}, g)$. Hence, the trajectory through the center of the dent yields the greatest effect. From these expressions we construct the solution for an incoming plane wave as
\begin{align}
    \chi_{\mathrm{t}}(x) =\ee^{\ii k \left( x_0 + s(x) \right)} \ ,
\end{align}
where $s(x)<0$ for $x<x_0$ and $s(x)\ge 0$ for $x\ge x_0$. We choose $x_0$ such that $x_0 \leq \mathrm{inf}\{\mathrm{supp}(f)\}$ or sufficiently small if $\mathrm{supp}(f)$ does not have a lower bound. Next, we can refer to the general ansatz
\begin{align}
    \chi_{\mathrm{t}}(x) = \begin{cases}
			\ee^{\ii k x} + a^{-} \ee^{-\ii k x} & : x \rightarrow -\infty \\
			  a^{+} \ee^{\ii k x} & : x \rightarrow +\infty	
		\end{cases} \ ,
\end{align}
to compute the one-dimensional analog of the scattering amplitude in the forward ($a^{+}$) and backward ($a^{-}$) directions. We find
\begin{align}
\begin{aligned}
 	a^{-} &= 0 \ , \\
	a^{+} &= \ee^{2\ii k \left( x_0 + s(0) \right)} - 1 \ .
\end{aligned}
\end{align}
Then, the one-dimensional analog of the total scattering cross length can be calculated as the discrete sum
\begin{align}
	\sigma_{\mathrm{tot}} = |a^{+}|^2 + |a^{-}|^2 &= \bigg|\ee^{2\ii k \left( x_0 + s(0)\right)} - 1 \bigg|^2 \notag \\
	& = | \ee^{\ii k \wp} - 1|^2 \ ,
\end{align}
identifying the effective path extension as 
\begin{align}
    \wp = \int_{-\infty}^{\infty} \left(\sqrt{1 + (\partial_{\tilde{x}} f)^2} - 1\right)\ \mathrm{d}\tilde{x}
\end{align}
since $x_0 \notin \mathrm{supp}(f)$ or sufficiently small. The last expression can be transformed to
\begin{align}
	\sigma_{\mathrm{tot}} = 2 \left[ 1 - \cos\left(k \wp \right) \right] \ , 
\end{align}
and the $k$-period of the scattering cross length $k_{\mathrm{p}}$ is given by
\begin{align}
	 k_{\mathrm{p}} = \frac{2 \pi}{\wp}\ .
\end{align}
This simplified consideration explains the periodicity of $\sigma_{\mathrm{tot}}$ in the wavenumber $k$ (see fig.\,\ref{fig:scat_data_w_delta}\,(l)) as an interference effect between the incoming wave and the scattered wave. The wavelength obtained from figs.\,\ref{fig:scat_data_w_delta}(k) and (l) deviates by approximately $6\%$ and $3\%$ from our analytical model.
\par
Furthermore, as the dent possesses a nonvanishing diameter that is given by the integrated cross length, it acts as an aperture. The geometric structure resembles a gravitational lens for light waves propagating through curved space. An alteration of the beam trajectory rationalizes both effects due to deformations of space (or spacetime) \citep{Grav_lens_Refsdal1964,Grav_lens_Bartelmann2001}. The difference here is that the extension of the curved region is comparable to the wavelength of the projectile so that a wave mechanical picture is justified. Note that, in contrast to electromagnetic radiation, we are dealing with wave functions $\chi_{\mathrm{t}}$ representing quantum particles, i.e., only the effects of the diffraction pattern on the observable $\vert \chi_{\mathrm{t}}\vert^{2}$ are physically meaningful. Regarding the diffraction pattern implies the idea of engineering nanostructures for particle-optical applications.

\section{Conclusions}
\label{sec:Conclusion}
We presented an analysis of the elastic channel for generalized quantum potential scattering confined to Riemannian manifolds. In addition to possible external scattering potentials, there exists an effective geometry-induced potential field that is always attractive when a two-dimensional subset of the three-dimensional Euclidean space is considered. However, this field is sensitive to geometric disturbances because of its nonlinear dependencies on the geometric parameters. Furthermore, the deviating metric yields contributions to the scattering amplitude. Analytical results and simulations (that have been extensively tested for convergence and reproduced accurately the known limiting cases such as flat space behavior) were presented and compared for a Gaussian dent within a flat two-dimensional space. The first Born approximation was found to be reliable only in the lower energy limit, where the geometric potential field is the dominant target. Particularly at high energies, the adapted metric tensor field leads to wave interference resulting in a diffraction pattern. This effect appears similar to a gravitational lens and is envisioned to be exploited in nanostructure engineering.
\par
As for experiments that can be studied with our formalism, we mention two-dimensional materials such as free-standing graphene or TMDC sheets. They develop ripples because long-range ordering in such systems of reduced dimension is unstable towards spatial fluctuations \cite{DENG2016197}. However, as the effects of metric modifications can be hard to distill in transport (scattering) experiments, since the wavelength of the transport electrons (the Fermi velocity is $v_{\mathrm{F}} \approx 3\cdot 10^{6} m/s$) is way smaller than the spatial variations, more promising examples for granting experimental access are (probably nanoscopic) tip-induced deformations. Indeed, a recent experiment indicates the formation of tip-induced bound states \cite{Harats2020}.

\appendix

\section{Derivation of the Scattering Amplitude}
The scattering problem is described by the Lippmann-Schwinger equation adapted to the effective tangent quantum dynamics with respect to a Riemannian manifold $\left( \mathcal{M}, g\right)$. Using cylindrical coordinates, the metric tensor field and the geometric potential field attain the forms
\begin{align*}
    \left(g_{\mu\nu} \right) &\simeq \left( \eta_{\mu\nu}\right) = \left( \begin{matrix}
    1 & 0 \\
    0 & r^{2}
    \end{matrix} \right)\ ,\ \left( r\rightarrow \infty \right), \nonumber\\
    \quad V_{\mathrm{geo}} &\simeq V_{\mathrm{geo}}^{0} = 0\ ,\ \left( r\rightarrow \infty \right)\ ,
\end{align*}
in the asymptotic region. The first Born approximation of the scattering amplitude \eqref{eq:scat_amp_LS} is performed by replacing $\vert \chi_{\mathrm{t}}^{+} \rangle \simeq \vert \chi_{\mathrm{t}}^{0}\rangle$. Therefore, the evaluation of the resulting integrand for the case of asymptotically flat manifolds that are parameterized in Monge form, so that \eqref{eq:Monge_i} -- \eqref{eq:Monge_f} hold, leads to
\begin{widetext}
\begin{align*}
    \langle\textbf{q}'\vert \hat{H}^{\mathrm{s}}\vert \chi_{\mathrm{t}}^{+} \rangle &\simeq \langle\textbf{q}'\vert \hat{H}^{\mathrm{s}}\vert \chi_{\mathrm{t}}^{0} \rangle \\
    &= \dfrac{1}{2\pi}\,\left[ -\dfrac{\hbar^{2}}{2m_{0}}\,\left( F^{2}\,\left( \dfrac{\textbf{k}\cdot \textbf{q}'}{r'}\right)^{2} - \ii\,\left( \dfrac{F^{2}}{r'} + F\,\partial_{r'}F \right)\,\dfrac{\textbf{k}\cdot \textbf{q}'}{r'} \right) + V_{\mathrm{geo}} \right]\,\ee^{\ii\textbf{k}\cdot\textbf{q}'} \\
    &= -\dfrac{\hbar^{2}}{4\pi m_{0}}\,\left[ F^{2}\,\left( \dfrac{\textbf{k}\cdot \textbf{q}'}{r'}\right)^{2} - \ii\,\left( \dfrac{F^{2}}{r'} + F\,\partial_{r'}F \right)\,\dfrac{\textbf{k}\cdot \textbf{q}'}{r'} + \dfrac{1}{4}\,\left( \dfrac{F}{r'} - \partial_{r'}F \right)^{2} \right]\,\ee^{\ii\textbf{k}\cdot\textbf{q}'}\ .
\end{align*}
Inserting this expression into \eqref{eq:scat_amp_LS} yields
\begin{align*}
    a(\textbf{k}, \textbf{k}^{\mathrm{s}}) &\simeq a^{(1)}(\textbf{k}, \textbf{k}^{\mathrm{s}}) \\
    &= -\dfrac{\ee^{-\ii\frac{3\pi}{4}}}{\sqrt{8\pi k}}\,\int \ee^{\ii\,\Delta\textbf{k}\cdot \textbf{q}'}\,\left[ F^{2}\,\left( \dfrac{\textbf{k}\cdot \textbf{q}'}{r'}\right)^{2} - \ii\,\left( \dfrac{F^{2}}{r'} + F\,\partial_{r'}F \right)\,\dfrac{\textbf{k}\cdot \textbf{q}'}{r'} + \dfrac{1}{4}\,\left( \dfrac{F}{r'} - \partial_{r'}F \right)^{2} \right]\ \mathrm{d}\sigma(\textbf{q}')\ ,
\end{align*}
\end{widetext}
where $\Delta\textbf{k}\coloneqq \textbf{k} - \textbf{k}^{\mathrm{s}}$ was introduced. Using it to define an internal $\tilde{x}$-axis and denoting the angle between the wave vectors as $\vartheta$, basic trigonometric considerations under incorporation of the conservation of momentum yield
\begin{align*}
    \Vert \Delta\textbf{k} \Vert_{2} &= 2\,k\,\mathrm{sin}\left( \sfrac{\vartheta}{2} \right) = 2\,k_{\tilde{x}} \\
    &\Rightarrow\ k_{\tilde{y}}^{2} - k_{\tilde{x}}^{2} = k^{2}\,\mathrm{cos}(\vartheta)\ .
\end{align*}
The above integral can be computed in polar coordinates using the Bessel integral of the first kind, and applying trigonometric relations, the result is expressed as function $\vartheta$ as
\begin{widetext}
\begin{align}
\begin{aligned}
    a^{(1)}(\vartheta) = \ee^{\ii\frac{\pi}{4}}\,\sqrt{\dfrac{\pi}{2k}}\,\int\limits_{0}^{\infty} r'\,&\left[ \left( F^{2}k^{2}\,\mathrm{sin}^{2}\left( \sfrac{\vartheta}{2} \right) + \dfrac{1}{4}\,\left( \dfrac{F}{r'} - \partial_{r'}F \right)^{2} \right)\,J_{0}(\rho') \right.\\
    & \hspace*{3mm}\left. + \left( \dfrac{F^{2} k}{2\,r'\,\mathrm{sin}\left( \sfrac{\vartheta}{2} \right)} + k\,\mathrm{sin}\left( \sfrac{\vartheta}{2} \right)\,F\,\partial_{r'}F \right)\,J_{1}(\rho') \right]\ \mathrm{d}r'\ .
\end{aligned}
\label{eq:1B_scat_amp}
\end{align}
Therein, $\rho' \coloneqq \Vert \Delta\textbf{k}\Vert_{2}\,r' = 2\,k\,r'\,\mathrm{sin}\left( \sfrac{\vartheta}{2} \right)$ and $J_{n}$ denotes the Bessel function of the first kind of order $n\in\mathbb{N}_{0}$. This expression gives the scattering amplitude in the fist Born approximation. In the main text, it is compared with the total scattering cross length obtained from numerical simulation.
\par
We note that the optical theorem \citep{Gu1989}
\begin{align}
    \sigma_{\mathrm{tot}} = 2\,\sqrt{\dfrac{\pi}{k}}\,\bigl( \Im\{a(0)\} - \Re\{a(0)\} \bigl) 
\end{align}
yields $\sigma_{\mathrm{tot}}^{(1)} \simeq 0$ when combined with the first Born approximation of the scattering amplitude \eqref{eq:1B_scat_amp}, because using the series expansion of the latter for small arguments we find that the forward scattering (for $\textbf{k}\parallel \textbf{k}^{\mathrm{s}}\ \Leftrightarrow\ \vartheta = 0$) is given by
\begin{align}
\begin{aligned}
    a^{(1)}(0) &= \ee^{\ii\frac{\pi}{4}}\,\sqrt{\dfrac{\pi}{32\,k}}\,\int\limits_{0}^{\infty} r'\,\left[ \left( \dfrac{F}{r'} - \partial_{r'}F \right)^{2} + 2\,k^{2}\,F^{2} \right]\ \mathrm{d}r' = \ee^{\ii\frac{\pi}{4}}\,\sqrt{\dfrac{\pi}{2k}}\,\int\limits_{0}^{\infty} r'\,\left[ -\dfrac{2 m}{\hbar^{2}}\,V_{\mathrm{geo}} + \dfrac{k^{2}\,F^{2}}{2} \right]\ \mathrm{d}r'\ . 
\end{aligned} 
\end{align}
\end{widetext}
This circumstance is not surprising as the relation is proven to be valid for the complete solution of the Lippman-Schwinger equation only and, obviously, breaks down during the Born approximation for axially symmetric asymptotically flat Riemannian manifolds.

\section{Derivation of the Expressions for PWA}
The usage of PWA to solve short-range potential scattering problems is well-documented in the literature. Here, we perform it for the sake of completeness and show its validity for scattering in asymptotically flat manifolds, i.\,e. for scattering upon spatially localized geometric perturbations.
\par
For axially symmetric manifolds parameterized by the radius $r$ from the symmetry center and the azimuthal angle $\vartheta$, the time-independent effective tangent Schrödinger equation, as derived from \eqref{eq:tangent_Schroedinger}, is cast as
\begin{align}
    \left[ \dfrac{v}{r}\,\partial_{r}\left( r\,v\,\partial_{r} \right) + \dfrac{1}{r^{2}}\,\partial_{\vartheta}^{2} + \left( k^{2} - U \right)\,\hat{I}\right] \chi_{\mathrm{t}}(r,\vartheta) = 0 
\label{eq:scat_prob_curv}
\end{align}
where we used $E= \frac{\hbar^{2}k^{2}}{2m_{0}}$. The reduced scattering potential is $U(r) \coloneqq \frac{2 m_{0}}{\hbar^{2}}\,V(r)$, and \linebreak$v(r) \coloneqq \frac{1}{\sqrt{1+(\partial_{r}f)^{2}(r)}}$. As the Hamilton operator $\hat{H}$ is invariant under rotations by the azimuthal angle $\vartheta$, $\hat{H}$ and the operator of axial orbital angular momentum $\hat{l}$ possess simultaneous eigenstates and the separation ansatz
\begin{align}
    \chi_{\mathrm{t}}^{0}(r,\vartheta) = \frac{v_{m}(r)}{\sqrt{r}}\,\ee^{\ii m\vartheta}\ ,\ m\in\mathbb{Z}
\label{eq:ansatz}
\end{align}
is appropriate. Thus, we obtain
\begin{align}
    \left[ \left( v\,\partial_{r} \right)^{2} + \left( k^{2} - \dfrac{m^{2}}{r^{2}} + \dfrac{v^{2}}{4r^{2}} - U_{\mathrm{eff}} \right)\,\hat{I} \right] u_{m}(r) = 0 
\label{eq:scat_prob_curv_ansatz}
\end{align}
with the effective reduced potential
\begin{align}
    U_{\mathrm{eff}} = U + \dfrac{1}{2}\,\dfrac{v\,\partial_{r}v}{r} = U - \dfrac{1}{2r}\,\dfrac{\partial_{r}f\,\partial_{r}^{2}f}{\sqrt{1+(\partial_{r}f)^{2}}}\ . 
\end{align}
In general, the scattering potential field is given as the sum of the geometric and external potential fields, \linebreak$V \doteq V_{\mathrm{geo}} + V_{\mathrm{ext}}$. For our parametrization, which gives \eqref{eq:Monge_i} -- \eqref{eq:Monge_f}, the case of free motion yields an effective reduced potential reading
\begin{align}
    U_{\mathrm{eff}} =&\, - \dfrac{1}{2r}\,\dfrac{\partial_{r}f\,\partial_{r}^{2}f}{\sqrt{1+(\partial_{r}f)^{2}}} - \dfrac{1}{4r^{2}}\,\dfrac{(\partial_{r}f)^{2}}{1+(\partial_{r}f)^{2}} \nonumber\\
    &\,+ \dfrac{1}{2r}\,\dfrac{\partial_{r}f\,\partial_{r}^{2}f}{\left( 1+(\partial_{r}f)^{2} \right)^{2}} - \dfrac{1}{4}\,\dfrac{(\partial_{r}^{2}f)^{2}}{\left( 1+(\partial_{r}f)^{2} \right)^{3}}\ , 
\label{eq:eff_red_pot}
\end{align}
explicitly. Setting $v(r) \doteq 1 + w(r)$, we can express \eqref{eq:scat_prob_curv_ansatz} as
\begin{widetext}
\begin{align}
    &\Biggl\{ \left[ \partial_{r}^{2} + \left( k^{2}r^{2} - m^{2} + \dfrac{1}{4} \right)\,\hat{I} \right] + \left[ \left( w^{2} + 2w \right)r^{2}\,\partial_{r}^{2} + \left( w + 1 \right)\partial_{r}w\,r^{2}\,\partial_{r} + \left( \dfrac{w^{2}}{4} + \dfrac{w}{2} - r^{2}U_{\mathrm{eff}} \right)\,\hat{I} \right]\Biggl\}\, u_{m}(r) = 0\ . 
\label{eq:scat_prob_full}
\end{align}
In contrast, in the asymptotic region equation \eqref{eq:scat_prob_curv} reduces to
\begin{align}
    \left[ \partial_{r}^{2} + \dfrac{1}{r}\,\partial_{r} + \dfrac{1}{r^{2}}\,\partial_{\vartheta}^{2} + \left( k^{2} - U^{0}\right)\,\hat{I} \right]\,\chi^{0}_{\mathrm{t}}(r,\vartheta) = 0\ , 
\label{eq:scat_prob_flat}
\end{align}
and inserting an ansatz similar to \eqref{eq:ansatz} yields
\begin{align}
    \left[ \partial_{r}^{2} + \left( \dfrac{1}{4r^{2}} + k^{2} - U^{0}(r) - \dfrac{m^{2}}{r^{2}} \right)\,\hat{I} \right]\,u^{0}_{m}(r) = 0 
\label{eq:scat_prob_asym}
\end{align}
as a remaining differential equation for the scaled radial part. Comparing \eqref{eq:scat_prob_asym} with \eqref{eq:scat_prob_full}, we see that
\begin{align*}
    \left[ \left( w^{2} + 2w \right)r^{2}\,\partial_{r}^{2} + \left( w + 1 \right)\partial_{r}w\,r^{2}\,\partial_{r} + \left( \dfrac{w^{2}}{4} + \dfrac{w}{2} - r^{2}U_{\mathrm{eff}} \right)\,\hat{I} \right] u_{m}(r) \overset{!}{\simeq} -U^{0}\,u_{m}^{0}(r) = 0\ ,\ (r\rightarrow \infty)
\end{align*}
\end{widetext}
must hold in the limit of large distances so that the solution of the differential equation converges appropriately to its asymptotic solution, i.\,e. $u_{m}(r) \simeq u_{m}^{0}(r)$, $r\rightarrow \infty$. As all functions $u_{m}$ and their derivatives have to be bounded due to the normalization condition of the wave function, we can extract the conditions
\begin{align*}
    w \in o(r^{-2})\ ,\qquad U_{\mathrm{eff}}\in o(r^{-2})\ .
\end{align*}
Henceforth, the vanishing $\partial_{r}f \rightarrow 0$ $(r\rightarrow \infty)$ is implied, because by truncating an expansion we find
\begin{align*}
    w(r) &= \dfrac{1}{\sqrt{1+(\partial_{r}f)^{2}}} - 1 \\
    &\simeq (\partial_{r}f)^{2}\ ,\ (r\rightarrow \infty)\ ,
\end{align*}
so that the following holds
\begin{align}
    w\in o(r^{-2})\ &\Leftrightarrow\ \partial_{r}f \in o(r^{-1})\nonumber\\
    &\Leftrightarrow\ f\in o(r^{-\alpha}) + c\ ,\ \alpha\in\mathbb{R}: \alpha > 0\ . 
\label{equiv:PWA_cond}
\end{align}
From \eqref{eq:eff_red_pot} we see that the condition concerning the effective reduced potential is automatically fulfilled when \eqref{equiv:PWA_cond} holds. Thus, the latter is a necessary and sufficient condition for applying the PWA. It states, that the manifold has to be a plane containing a dent that decays at least with a power law.
\par
While \eqref{eq:scat_prob_curv} describes a scattering problem within curved two-dimensional space, \eqref{eq:scat_prob_flat} enables the consideration of scattering upon the entire curved region. As we deal with the same ansatz that was used by Lapidus \citep{Lapidus1982} for considerations of potential scattering in flat two-dimensional spaces, the same results are valid for our situation of scattering upon curved regions. From the requirement of regularity at the center, i.\,e. for $r\rightarrow 0$, and the short range characteristics of $U$, we deduce the asymptotic behavior of the solutions
\begin{align}
\begin{aligned}
    u_{m}(r) &\overset{r\rightarrow 0}{\propto} r^{m + \frac{1}{2}}\ ,\\
    u_{m}(r) &\overset{r\rightarrow\infty}{\propto} s_{m}(kr) - \mathrm{tan}(\delta_{m})\,c_{m}(kr)\ ,
\end{aligned} 
\label{eq:vl_asy}
\end{align}
where $\Theta_{m}(kr) \coloneqq kr-\frac{\pi}{2}\left( m + \frac{1}{2} \right)$ (\citep{Standards2010, Lapidus1982, Burke2011})
\begin{align*}
    s_{m}(kr) &\propto \sqrt{kr}\,J_{m}(kr) \overset{r\rightarrow\infty}{\propto} \sqrt{\dfrac{2}{\pi}}\,\mathrm{cos}(\Theta_{m}(kr))\ ,\\
    c_{m}(kr) &\propto \sqrt{kr}\,Y_{m}(kr) \overset{r\rightarrow\infty}{\propto} \sqrt{\dfrac{2}{\pi}}\,\mathrm{sin}(\Theta_{m}(kr))
\end{align*}
are the regular and singular solutions in dependence of the Bessel functions of the first and second kind, respectively, and $\delta_{m}$ is the scattering phase shift of the corresponding orbital angular momentum channel. Choosing $N_{m} = 2\,\mathrm{cos}(\delta_{m})\,\ee^{\ii\delta_{m}}$ as the respective normalization constant within \eqref{eq:vl_asy} gives the equivalent expression
\begin{align*}
    u_{m}(r) &\overset{r\rightarrow\infty}{\propto} \sqrt{kr}\,\left( S_{m}\,H_{m}^{(1)}(kr) + H_{m}^{(2)}(kr) \right) \\ &\overset{r\rightarrow\infty}{\propto} \sqrt{\dfrac{2}{\pi}}\,\left( \ee^{-\ii\Theta_{m}(kr)} + S_{m}\,\ee^{\ii\Theta_{m}(kr)} \right)\ ,
\end{align*}
that includes the Hankel functions of first and second kind and the S-matrix $S_{m} \coloneqq \ee^{2\ii\delta_{m}}$. Adapting the reasoning of Burke \citep{Burke2011}, we find
\begin{align}
    K_{m} = \mathrm{tan}(\delta_m) = -\dfrac{\pi}{2}\,\dfrac{1}{\sqrt{k}}\,\int\limits_{0}^{\infty} \sqrt{r}\,U(r)\,u_{m}(r)\,J_{m}(kr)\ \mathrm{d}r 
\label{eq:KMatrix}
\end{align}
as an explicit expression for the $K$-matrix.
\par
As for the superposition ansatz \eqref{eq:wave_profile}, we represent the incoming plane wave according to the Jacobi-Anger expansion \citep{Standards2010}
\begin{align*}
    \ee^{\ii kr\,\mathrm{cos}(\vartheta)} = \sum\limits_{m=-\infty}^{+\infty} \ii^{m}\,J_{m}(kr)\,\ee^{\ii m\vartheta}\ ,
\end{align*}
and, similarly, expand the entire wave formally with respect to the same set of basis functions,
\begin{align*}
    \chi_{\mathrm{t}}(r,\vartheta ) = \sum\limits_{m=-\infty}^{+\infty} B_{m}(k)\,\dfrac{u_{m}(kr)}{\sqrt{r}}\,\ee^{\ii m\vartheta}\, ,\ B_{m}\in\mathbb{C}\ \forall\ m\in\mathbb{Z} \,.
\end{align*}
Inserting these terms in \eqref{eq:wave_profile} and considering the asymptotic forms, we find
\begin{widetext}
\begin{align*}
    &\sum\limits_{m=-\infty}^{+\infty} \dfrac{B_{m}(k)}{2\,\mathrm{cos}(\delta_{m})}\,\sqrt{\dfrac{2}{\pi r}}\,\left( \ee^{\ii\,(\Theta_{m} + \delta_{m})} + \ee^{-\ii\,(\Theta_{m} + \delta_{m})} \right)\,\ee^{\ii m\vartheta}= \dfrac{1}{2\pi}\, \left( \sum\limits_{m= -\infty}^{+\infty} \dfrac{\ii^{m}}{2}\,\sqrt{\dfrac{2}{\pi kr}}\,\left( \ee^{\ii\Theta_{m}} + \ee^{-\ii\Theta_{m}} \right)\,\ee^{\ii m\vartheta} + a(\vartheta)\,\dfrac{\ee^{\ii kr}}{\sqrt{r}} \right)\ ,
\end{align*}
\end{widetext}
From this we deduce
\begin{align*}
    B_{m}(k) = \ii^{m}\,\dfrac{\mathrm{cos}(\delta_{m})}{2\pi\,\sqrt{k}}\,\ee^{\ii\delta_{m}}
\end{align*}
by a comparison of coefficients for the expansions according to orthogonal basis functions and find the identity
\begin{align}
    a(\vartheta) = \dfrac{\ee^{-\ii\frac{\pi}{4}}}{\sqrt{2\pi k}}\,\sum\limits_{m=-\infty}^{+\infty}\left( S_{m} - 1 \right)\,\ee^{\ii m\vartheta}\ . 
\end{align}
This is the relation we state in the main text. Inserting it into the definitions of the scattering cross lengths, \eqref{eq:total_cross_length} and \eqref{eq:momtrans_cross_length}, yields the expressions \eqref{eq:total_cross_length_PWA} and \eqref{eq:momtrans_cross_length_PWA}, respectively. Furthermore, we are able to express the orbital angular momentum of the scattered wave with respect to the $z$-axis as
\begin{align}
    L^{\mathrm{s}} = \dfrac{\hbar}{\sigma_{\mathrm{tot}}}\,\dfrac{1}{k}\,\sum\limits_{m=-\infty}^{+\infty} m\,\vert S_{m} - 1 \vert^{2}\ ,
\end{align}
but as $S_{m} = S_{-m}\ \forall\ m\in\mathbb{Z}$ for an axially symmetric scattering potential we expect vanishing orbital angular momentum transfer in our case. Finally, if the solution $\chi_{\mathrm{t}}$ is known upon $\mathrm{supp}(U)$, we can find an alternative expression of the the K-matrix \eqref{eq:KMatrix} by equating the ratios between radial projections of the wave function and its values according to \eqref{eq:vl_asy} taken at two distinct but sufficiently large radii, and isolating the tangent of the scattering phase shift. The result is given by the formula \eqref{eq:K_mat_eval} in the main text.

\section{Discussion of the Geometric Potential}
\begin{figure*}[!]
\begin{overpic}[width=0.67\textwidth]{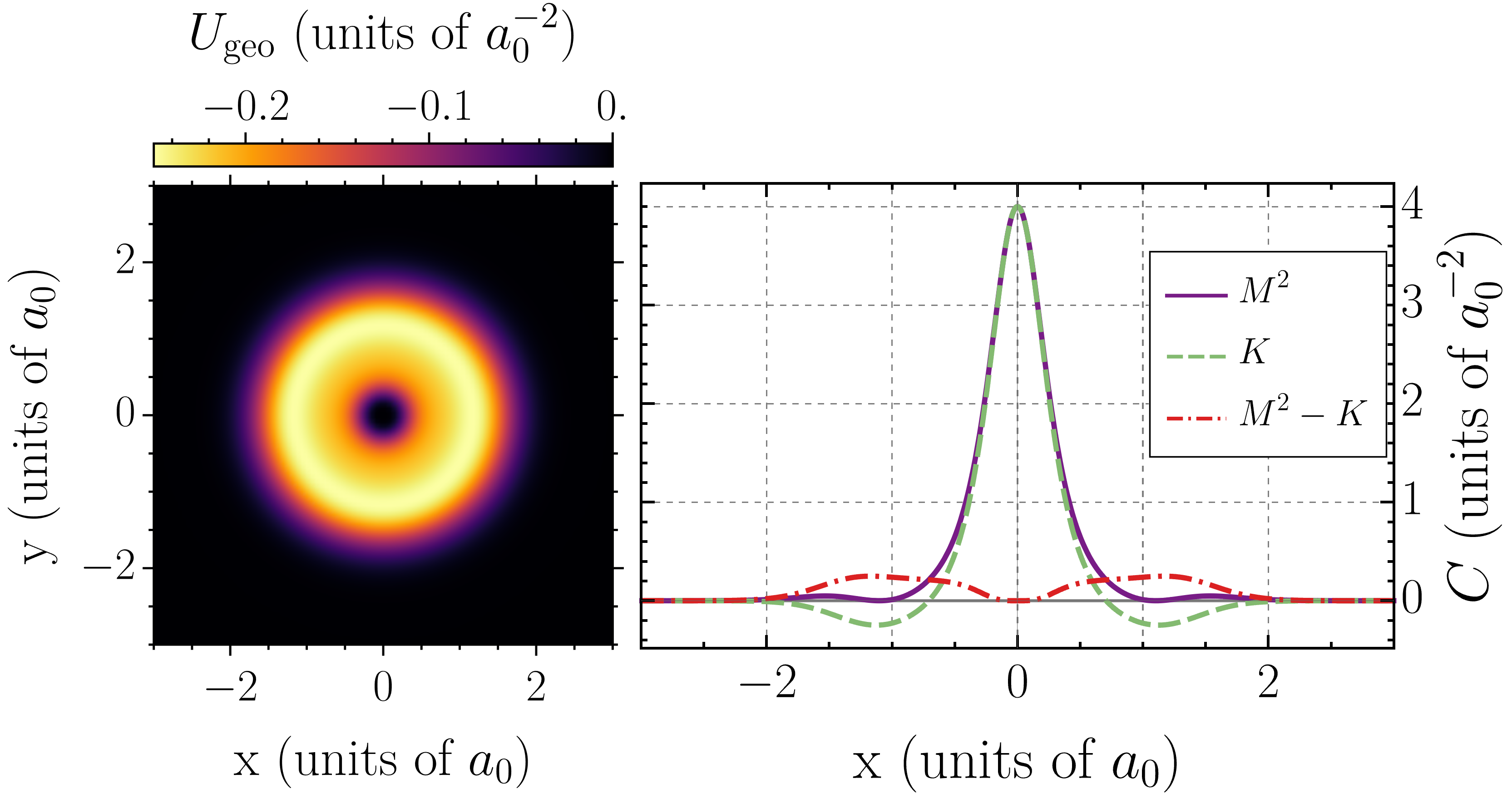}
\put(1,40){(a)}
\put(96,40){(b)}
\end{overpic}
\centering
\caption{(a) Density plot of the reduced geometric potential $U_{\mathrm{geo}}$ within the parameter space and (b) course of the Gaussian and mean curvatures $K$ and $M$ for a Gaussian dent with parameters $f_{0} = 1\,a_{0}$ and $\sigma = \frac{1}{\sqrt{2}}\,a_{0}$, with $a_{0}$ describing an arbitrary length unit. The geometric potential consists of two contributions dependent on respectively one type of curvature, and their difference yields an axially symmetric ring-shaped structure with radial dependence.}
\label{fig:Ugeo_curv}
\end{figure*}
Considering the Gaussian dent as given by \eqref{eq:Gauss_Monge}, the relevant quantities of differential geometry can be evaluated explicitly. The covariant metric tensor field \eqref{eq:Monge_i} has the matrix representation
\begin{align*}
    \left( g_{\mu\nu} \right) = \left( \begin{matrix}
    \left[ 1 + \left( \frac{f_{0}\,r}{\sigma^{2}} \right)^{2}\,\ee^{-\frac{r^{2}}{\sigma^{2}}} \right] & 0 \\
    0 & r^{2}
\end{matrix}\right) 
\end{align*}
in view of the natural local basis. With the expression $F$ given in \eqref{eq:Gauss_F} we find the Gaussian and mean curvatures according to \eqref{eq:Monge_pot},
\begin{align*}
    K &= \dfrac{f_{0}^{2}\,\sigma^{2}\,\left( \sigma - r \right)\,\left( \sigma + r \right)\,\ee^{\frac{r^{2}}{\sigma^{2}}}}{\left( f_{0}^{2}\,r^{2} + \sigma^{4}\,\ee^{\frac{r^{2}}{\sigma^{2}}} \right)^{2}}\ ,\\
    M &= \dfrac{f_{0}\,\left( -f_{0}^{2}\,r^{2} + \sigma^{2}\,\left( r^{2} - 2\sigma^{2} \right)\,\ee^{\frac{r^{2}}{\sigma^{2}}} \right)\,\ee^{-\frac{r^{2}}{2\sigma^{2}}}}{2\,\sigma^{2}\,\left( f_{0}^{2}\,r^{2} + \sigma^{4}\,\ee^{\frac{r^{2}}{\sigma^{2}}} \right)\,\sqrt{1 + \left( \frac{f_{0}\,r}{\sigma^{2}} \right)^{2}\,\ee^{-\frac{r^{2}}{\sigma^{2}}}}}\ ,
\end{align*}
and derived the geometric potential \eqref{eq:Gauss_Vgeo} from them. The result for the reduced potential $U_{\mathrm{geo}}\coloneqq \frac{2\,m_{0}}{\hbar^{2}}\,V_{\mathrm{geo}}$ is shown in fig.\,\ref{fig:Ugeo_curv}\,(a) and exhibits an axially symmetric, ring-shaped structure as all the above mentioned quantities are functions of the radial variable $r$. The potential vanishes at the center (i.\,e. $r\rightarrow 0$) where $K(0)=\frac{f_{0}^{2}}{\sigma^{4}}$ and $M(0) = -\frac{f_{0}}{\sigma^{2}}$. As can be seen from the equations, the geometric potential is nonlinearly dependent on both the amplitude $f_{0}$ and the width measure $\sigma$. Hence, narrowing the width or increasing the amplitude results in larger curvatures. However, the plots in fig.\,\ref{fig:Ugeo_curv}\,(b) indicate that the structure of the geometric potential can not be identified from or associated with either of the curvatures alone. For the Gaussian dent the two types of curvature yield quite similar contributions to the geometric potential so that their difference has a significantly lower range and a different shape. Especially, the potential vanishes where the curvatures reach their extremes. Nevertheless, we tested that a variation of the dent in the mentioned way leads to a stronger geometric potential, while its ring-shaped structure stays invariant.
\par
If we assume $a_{0}=2\,\mathrm{nm}$ which seems experimentally feasible in current 2D materials, we find amplitude values of $\approx 2.38\,\mathrm{meV}$. Therefore, the geometry-induced potential is significant for low projectile energies only. The varying metric tensor accompanying its occurrence still has an effect that depends on the ratio between the characteristic size of the geometric structure and the wavelength of the projectile.
\par
Due to symmetry, it is expected that any influence on an incident plane wave will be mirror symmetric with respect to the propagation direction defined by $\textbf{k}$. Therefore, only linear momentum but no angular momentum transfer from the projectile to the target should occur.

\FloatBarrier
\bibliography{scatteringI}

\end{document}